\documentstyle[12pt,epsf]{article}

\newcounter{saveeqn}
\newcounter{App} 

\newcommand{\gapprox}{\raisebox{-.2ex}{$\stackrel{\textstyle>}
{\raisebox{-.6ex}[0ex][0ex]{$\sim$}}$}}
\newcommand{\lapprox}{\raisebox{-.2ex}{$\stackrel{\textstyle<}
{\raisebox{-.6ex}[0ex][0ex]{$\sim$}}$}}

\begin{document}

\begin{flushright}
CEBAF-TH-95-11
\end{flushright}
\vspace{2cm}
\begin{center}
{\bf Quark-Hadron Duality and $\gamma^* p \to \Delta$ Form Factors}
\end{center}
\begin{center}{V.M. Belyaev$^*$ }
\end{center}
 \begin{center}
 {\em Continuous Electron Beam Accelerator Facility  \\
 Newport News, VA 23606, USA}
\end{center}
\begin{center}{A.V. Radyushkin}  \\
{\em Physics Department, Old Dominion University,}
\\{\em Norfolk, VA 23529, USA}
 \\ {\em and} \\
{\em Continuous Electron Beam Accelerator Facility,} \\
 {\em Newport News,VA 23606, USA}
\end{center}
\vspace{2cm}
\begin{abstract}
We use  local quark-hadron duality
to estimate the  purely
nonperturbative soft contribution
 to the $\gamma^* p \to \Delta$ form factors.
 Our  results are in   agreement with  existing
experimental data.
We predict that the ratio
$G_E^*(Q^2)/G_M^*(Q^2)$ is small
for all accessible $Q^2$, in contrast to the
pQCD expectations that  $G_E^*(Q^2) \to -G_M^*(Q^2)$.
\end{abstract}
\vspace{1cm}
\flushbottom{$^*\overline{On\;  leave\;  of\;  absence }
\;from\;  ITEP,\;  117259\;  Moscow,\; Russia.$}

\newpage

\section{Introduction}

Basically, there are
two competing explanations
of the
experimentally observed power-law behaviour
of elastic hadronic form factors:
hard scattering \cite{bf} and the Feynman mechanism \cite{feynman}.
At sufficiently  large momentum transfer,
the Feynman mechanism  contribution is dominated
by  configurations  in which  one of
the quarks carries  almost
all the momentum of the hadron.
In QCD,  this results in  an extra  $1/Q^2$ suppression
compared to  the hard scattering term
generated by  the
valence configurations with  small
transverse sizes and  finite light-cone fractions
of the  total hadron momentum
carried by each valence  quark.
The hard term, which   eventually  dominates,
 can be written in a  factorized form \cite{cz},\cite{er},\cite{lb},
 as a product of the
perturbatively calculable hard scattering amplitude and
two distribution amplitudes accumulating
the necessary nonperturbative information.
However, this mechanism involves exchange
of virtual gluons, each exchange
 bringing in a suppression factor $(\alpha_s/{\pi}) \sim  0.1$.
Hence, to describe  existing data
by the hard contribution alone,
one should intentionally  increase
the magnitude of the hard scattering term by using
distribution amplitudes
with  a peculiar ``humped''  profile \cite{cz2}.
In this case,   passive quarks
carry  a small fraction of the hadron momentum and,
 as   pointed out in ref.\cite{isgur},  the ``hard'' scattering
subprocess,   even at  rather  large momentum transfers
$Q^2\sim 10\, GeV^2$, is dominated, in fact,  by very small  gluon
 virtualities.
This means that the hard scattering scenario heavily relies
on the assumption that the asymptotic pQCD
expressions  are accurate even for
momenta smaller than  $300 \, MeV$,
$i.e.,$ in the region strongly affected by
finite-size effects, nonperturbative QCD vacuum fluctuations
$etc.$   Including  these effects
shifts   the hard contributions significantly
below the data level even if one uses
the  humpy distribution amplitudes
and other {\it ad hoc} modifications intended to
increase the hard term (see, $e.g.,$ \cite{kroll}).

Furthermore, as argued in ref.\cite{mr},
the derivation of the humpy distribution
amplitudes in \cite{cz2} implies a
rather singular  picture (infinite correlation length)
of the QCD vacuum fluctuations.
Under  realistic assumptions,
it is impossible to get
distribution amplitudes strongly differing from the
smooth ``asymptotic'' forms. The latter are known to produce
hard contributions which  are too small to describe the data
on elastic form factors.
Thus, there is an increasing evidence in favour of
the alternative  scenario, $viz.,$  that
for  experimentally accessible momentum transfers
the form factors are
still dominated  by the purely
soft contribution  corresponding  to the Feynman mechanism.

In the language of the light-cone formalism \cite{lb},
the soft term is given by  the overlap of
the soft parts of the  hadronic wave functions,
$i.e.,$ is an essentially nonperturbative object.
Among the existing approaches to the nonperturbative effects in QCD,
that which is closest to pQCD is  the QCD sum rule method \cite{SVZ}.
QCD sum rules were originally used to calculate
the soft contribution
for the pion form factor
in the region of moderately large
\cite{i1},\cite{nr82} and  then
small momentum transfers
 \cite{nr84}.
It should be emphasized that,
in the whole region $0  \leq Q^2  \lapprox 3 \, GeV^2$,
the results are very close to the experimental data:
the Feynman mechanism alone is sufficient to explain
the observed behaviour of the pion form factor.
 For higher $Q^2$,  the direct QCD sum rule method
fails   due to increasing  contributions
from higher condensates.   However,
a model summation of the higher terms
 into nonlocal condensates  \cite{mr} indicates
 that the soft term
dominates up to  $Q^2  \sim 10 \, GeV^2$ \cite{br}.
This conclusion is  also  supported by a  recent
calculation within    the framework of
the light-cone  sum rules \cite{braunhal}.

An important  observation made in ref.\cite{nr82}
is that the results of the  elaborate
QCD sum rule analysis are rather accurately
reproduced by  a  simple local quark-hadron
duality prescription.
The latter states  that
one can get an estimate for a hadronic   form factor
by considering  transitions between the
free-quark states produced  by a local current
having the  proper quantum numbers,
with subsequent  averaging  of
the invariant mass of the quark states
over the appropriate duality interval $s_0$.
The duality interval has a specific value for each
hadron, $e.g.,$ $s_0^{\pi} \approx 0.7 \, GeV^2$
for the pion and  $s_0^{N} \approx 2.3 \, GeV^2$ for the nucleon.

The local duality  ansatz,
equivalent to fixing
the  form of a soft wave function,
was used
to estimate  the soft contribution in the case of  the
proton  magnetic form factor \cite{nr83}.
The results agree with available data \cite{slac36},
\cite{slac883} over a wide region,
$3 \, GeV^2 \lapprox Q^2 \, \lapprox \,  20\, GeV^2$.
Furthermore, the  calculation
of ref.\cite{nr83}
 correctly reproduces (without any adjustable parameter),
the observed magnitude
 of the helicity-nonconservation  effects
$ F_2^p(Q^2)/F_1^p(Q^2) \sim \mu^2/Q^2 $
with $\mu^2 \sim 1 \, GeV^2 $ \cite{slac883}.
It is  difficult to understand
the origin of such a large scale
 within the hard scattering scenario,
since possible sources of  helicity nonconservation
 in pQCD include
only small scales like quark masses,
intrinsic transverse momenta $etc.$, and
one would rather expect that
$ \mu^2 \sim 0.1 \, GeV^2$.
 Thus,  the study of   spin-related
properties provides  a promising way
for an unambiguous  discrimination between
soft and hard scenarios.

Of a particular interest there is
the  $\gamma^* p \to \Delta$ transition.
A special attention to this process was
raised by the results \cite{stoler1} of the analysis of inclusive
SLAC  data which indicated that the relevant
form factor drops faster than predicted by
the  quark counting rules.
The  relevant hard scattering contribution
was originally considered in ref.\cite{carlson},
where it was observed  that
the hard scattering amplitude in this case
has an extra suppression due to cancellation
between  symmetric and antisymmetric parts
of the nucleon distribution amplitude,
and it was conjectured  that the faster fall-off
found in \cite{stoler1} can be explained by  the dominance
of some non-asymptotic  contribution.
Later, it was claimed \cite{stefber} that,
by appropriately choosing the distribution amplitudes,
one can get a leading-twist hard term
comparable in magnitude with  the data.
Furthermore, the results of
a recent reanalysis \cite{keppel} of the inclusive SLAC data
are rather consistent with the  $1/Q^4$
behaviour, and this revived the hope that
the $\gamma^* p \to \Delta$ form factor can be  still
described by pQCD.

However, the important result of the pQCD calculation  \cite{carlson}
is that the lowest-twist hard
 contribution has the property
$G_E^{* \, hard}(Q^2) \approx - G_M^{* \, hard} (Q^2)$.
Experimentally, the ratio $G_E^*(Q^2)/G_M^*(Q^2)$
is rather small \cite{burkert,burkert95}, which indicates that
the leading-twist pQCD term is irrelevant
in the region $Q^2 \lapprox \, 3 \, GeV^2$.
In the present paper,   we use the local quark-hadron duality
 to   estimate the soft contribution
for the $G_E^*(Q^2)$ and $G_M^*(Q^2)$
form factors of the $\gamma^* p \to \Delta$ transition
to  study whether the soft
contribution  is  large  enough to
describe the data and whether the relative smallness
of the electric form factor persists
in the  region  of moderately large
momentum transfers $3 \lapprox Q^2 \lapprox \, 15 \, GeV^2$.

\section{Three-point function and form factors}

The starting   object for a QCD sum rule analysis of the
$\gamma^* p \to \Delta$ transition is the 3-point correlator:
\begin{eqnarray}
T_{\mu\nu}(p,q)=\int
\langle 0|T\{ \eta_\mu(x) J_\nu (y) \bar{\eta}(0)\} |0\rangle
e^{ipx-iqy} d^4x d^4y
\label{1}
\end{eqnarray}
of  the electromagnetic current
\begin{eqnarray}
J_\nu=e_u\bar{u}\gamma_\nu u+e_d\bar{d}\gamma_\nu d
\label{2}
\end{eqnarray}
and two Ioffe currents  \cite{ioffe}
\begin{eqnarray}
\eta = \varepsilon^{abc} \left(u^a{\cal C}\gamma_\rho
u^b\right)\gamma_\rho\gamma_5d^c \ ,
\ \ \
\eta_\mu=\varepsilon ^{abc} \left( 2 \left( u^a {\cal C}
\gamma_\mu d^b \right) u^c+\left(u^a{\cal C}
\gamma_\mu u^b\right)d^c\right) \ .
\label{3}
\end{eqnarray}

We use the following parameterization for  the projections of
$\eta$ and  $\eta_\mu$ onto the
nucleon and $\Delta$-isobar states, respectively:
\begin{eqnarray}
\langle 0|\eta |N\rangle =\frac{ l_N }{(2\pi)^2} v \ ,
 \  \ \  \
\langle 0|\eta_\mu|\Delta\rangle =\frac{l_\Delta}{(2\pi)^2} \psi_\mu \ .
\label{4}
\end{eqnarray}
Here, $v$ is the Dirac spinor of the nucleon
while $\psi_\mu$ is the spin-3/2  Rarita-Schwinger wave function
for the $\Delta$-isobar, $i.e.,$
$(\hat{p}-\hat{q}-m)v=0$,
$(\hat{p}-M)\psi_\mu=0$,  \,
$p_\mu \psi_\mu=0$, $\gamma_\mu \psi_\mu=0$;
with $m$ being the nucleon mass and $M$ that of
$\Delta$.  We use the
notation $\hat{a} \equiv a_\alpha \gamma_\alpha$.

On  the hadronic level,
the $\gamma^* p \to \Delta$ transition  makes the following
contribution to
the correlator (\ref{1}):
\begin{eqnarray}
T_{\mu\nu}^{\gamma^* p \to \Delta}
=  \frac{  l_N l_\Delta}{(2 \pi)^4}
{ {X_{\mu \alpha}(p)}\over{p^2-M^2} } \,
\Gamma_{\alpha\nu} (p,q) \gamma_5 \,
{{\hat{p}-\hat{q}+m}\over{(p-q)^2-m^2}} \ ,
\label{5}
\end{eqnarray}
where $ \Gamma_{\alpha\nu} (p,q) \gamma_5$ is the $\gamma^* p \to \Delta$
vertex function
\begin{eqnarray}
\Gamma_{\alpha\nu}(p,q)=
G_1(q^2) \left (q_\alpha\gamma_\nu- g_{\alpha\nu} \hat{q} \right )
+
G_2(q^2)\left (q_\alpha P_\nu - g_{\alpha\nu} (qP) \right )
\nonumber
\\
+G_3 (q^2) \left (q_\alpha q_\nu- g_{\alpha\nu} q^2 \right )
\label{6}
\end{eqnarray}
($ P \equiv p - q/2$ ) and $X_{\mu \alpha}(p)$
the projector onto the isobar state
\begin{eqnarray}
X_{\mu \alpha}(p)=
\left(g_{\mu\alpha}-\frac13\gamma_\mu\gamma_\alpha
 +\frac1{3M}(p_\mu\gamma_\alpha-p_\alpha\gamma_\mu)-\frac{2}{3M^2}
p_\mu p_\alpha\right)(\hat{p}+M).
\label{7}
\end{eqnarray}

The form factors $G_1, G_2,G_3$ are related to
a more convenient set $G_E^*, G_M^*, G_C^*$ by
\begin{eqnarray}
G_M^*(Q^2)=\frac{m}{3(M+m)}\left(
((3M+m)(M+m)+Q^2)\frac{G_1(Q^2)}{M}\right.
\nonumber
\\
\left. +(M^2-m^2)G_2(Q^2)-2Q^2G_3(Q^2)
\right) \ ,
\label{8}
\end{eqnarray}
\begin{eqnarray}
G_E^*(Q^2)=\frac{m}{3(M+m)}\left(
(M^2-m^2-Q^2)\frac{G_1(Q^2)}{M}\right.
\nonumber
\\
\left. +
(M^2-m^2)G_2(Q^2)-2Q^2G_3(Q^2)\right) \ ,
\label{9}
\end{eqnarray}
\begin{eqnarray}
G_C^*(Q^2)=\frac{2m}{3(M+m)}\left(
2M G_1(Q^2)
+\frac12(3M^2+m^2+Q^2)G_2(Q^2) \right.
\nonumber
\\
\left.
+ (M^2-m^2-Q^2)G_3(Q^2)
\right) \ ,
\label{10}
\end{eqnarray}
(see, $e.g.,$ \cite{scadron}). One should not confuse
the  magnetic form factor $G_M^*(Q^2)$ given by eq.(\ref{8})
with the effective form factor
mentioned in refs.\cite{carlson},\cite{stoler}.
In particular, the form factor $G_T(Q^2)$ defined
by eq.(6.2) of ref.\cite{stoler}
can be written in terms of $G_M^*(Q^2)$ and  $G_E^*(Q^2)$
(defined by eqs.(8),(9) above)  as
\begin{equation}
|G_M^*|^2 + 3 |G_E^*|^2  = \frac{Q^2}{Q^2+ \nu^2}
\left ( 1 + \frac{Q^2}{(M+m)^2} \right )
 |G_T|^2  \ ,
\label{GT}
\end{equation}
where $\nu = (M^2 -m^2 +Q^2)/2m$ is the energy of the
virtual photon in the proton rest frame.
Note  that, for large $Q^2$, our  $G_M^*(Q^2)$ and $G_T(Q^2)$ of
eq.(6.2) of ref.\cite{stoler} have the same power  behaviour.

\section{Local quark-hadron duality}

Multiplying all the factors in eq.(\ref{5}) explicitly,
one ends up with  a rather long sum of different
structures  $a^i_{\mu\nu}$
accompanied by the relevant invariant
amplitudes  $T_i$, each of which is a  combination
of the three independent transition form factors listed above.
To incorporate the  local quark-hadron duality, we write down the
dispersion relation
for each of the invariant amplitudes:
\begin{eqnarray}
T_i(p_1^2,p_2^2,Q^2)=
\frac1{\pi^2}\int_0^\infty ds_1\int_0^\infty ds_2
\frac{\rho_i(s_1,s_2,Q^2)}{(s_1-p_1^2)(s_2-p_2^2)}+
``subtractions" \ ,
\label{11}
\end{eqnarray}
where $p_1^2=(p-q)^2$, $p_2^2=p^2$. The perturbative contributions
to the amplitudes $T_i(p_1^2,p_2^2,Q^2)$ can also be written in the form
of eq.(\ref{11}). Evidently,
 the physical perturbative spectral densities
$\rho_i(s_1,s_2,Q^2)$ differ from their
perturbative analogues, the difference being most pronounced
in the resonance region,
$i.e.,$ for small $s_1$ and $s_2$ values. In particular,
$\rho_i(s_1,s_2,Q^2)$ contains the double
 $\delta$-function term corresponding
to the $\gamma^* p \to \Delta$ transition:
\begin{eqnarray}
\rho_i(s_1,s_2,Q^2) \sim
 l_N l_{\Delta}
F_i(s_1,s_2,Q^2)  \delta(s_1-m^2)\delta(s_2-M^2) \ ,
\label{12}
\end{eqnarray}
while the perturbative
spectral densities $\rho_i^{pert}(s_1,s_2,Q^2)$
are smooth functions of $s_1$ and $s_2$.
The local quark-hadron duality assumes, however,
that the two spectral densities
are  in fact dual to each other:
\begin{eqnarray}
\int_0^{s_0}ds_1\int_0^{S_0}\rho_i^{pert.}(s_1,s_2,Q^2) \, ds_2
=
\int_0^{s_0}ds_1\int_0^{S_0} \rho_i(s_1,s_2,Q^2) \, ds_2 \ ,
\label{13}
\end{eqnarray}
$i.e.,$  they give the same result
when integrated over the appropriate duality
 intervals $s_0,S_0$. The latter
characterize the effective thresholds for
higher states with the nucleon or, respectively,
 $\Delta$-isobar quantum numbers.
As noted in ref.\cite{nr83}, the local duality prescription  can be
treated as a model for the soft wave functions:
$$\Psi_N(\{x\},\{k_{\perp}\}) \sim
\theta \left ( \sum_{i=1}^3 \frac{k_{\perp_i}^2}{x_i} \leq s_0 \right )
\, ; \  \Psi_{\Delta}(\{x\},\{k_{\perp}\}) \sim
\theta \left (\sum_{i=1}^3 \frac{k_{\perp_i}^2}
{x_i} \leq S_0 \right )\, , $$
$i.e.,$ $s_0$ and $S_0$  also set the scale for the
width of the transverse momentum distribution.
Using this model, we can obtain
a good  estimate for the overlap
of the soft wave functions only in the intermediate $Q^2$-region
where the soft contribution is sensitive
 mainly to the  $k_{\perp}$-widths of the quark distributions
rather than to their detailed forms.
{}From  experience with the proton form factor calculations,
we expect that local duality  will work
in the region between  $3 \, GeV^2$ and $20 \, GeV^2$.
The low-$Q^2$ region $Q^2 \lapprox 1 \, GeV^2$,
in which there appear  large nonperturbative
contributions due to the long-distance
propagation in the $Q^2$-channel,  can  be
analyzed within a full-framed QCD sum
rule approach supplemented  by the formalism of
induced condensates \cite{induced} or
 bilocal operators \cite{bil}.

Applying the local quark-hadron duality
to the two-point correlators of
 $\eta$-  or, respectively, $\eta_{\mu}$-currents,
we obtain  simple relations between the
duality intervals $s_0$, $S_0$ and the
residues $l_N$,  $ l_\Delta$ of the Ioffe currents:
\begin{equation}
 l_N^2=\frac{s_0^3}{12}
\ \  \   \   \  ;  \  \  \   \  \
  l_\Delta^2=\frac{S_0^3}{10} \ .
\label{14}
\end{equation}
After the duality intervals are fixed
($e.g.,$ from the QCD sum rule analysis of the relevant
two-point function  \cite{belioffe}),
the local duality  estimates for the form factors
do not have any free parameters.

\section{Invariant amplitudes}

Now, choosing a particular Lorentz structure $a^i_{\mu\nu}$,
one can get the local duality estimate for the relevant
combination of the form factors.
One should remember, however, that
not  all the invariant amplitudes are equally reliable.
To compare the contributions related to different
structures, one should specify a reference frame.
In our case, the most relevant  is the infinite
momentum frame where
$p^{\mu} \equiv p^{\mu}_{\|} \to \infty$
while $q^{\mu} \equiv q^{\mu}_{\perp} $ is fixed.
So, $a \ priori$, the structures with the maximal number of the
``large'' factors $p^{\mu}$ are  more  reliable
than  those in which $p^{\mu}$  is  substituted by the ``small''
parameter $q^{\mu}$ or by $g_{\mu \nu}$.
However,  dealing with the $\eta_{\mu}$-current
in the $\Delta$-channel, one should realize
that $\eta_\mu$ has also a nonzero  projection
onto the spin-1/2 isospin-$3/2$
states $|\Delta^*(p)\rangle $:
\begin{eqnarray}
\langle 0|\eta_\mu|\Delta^*(p)\rangle =
\lambda^*  (m \gamma_\mu-4 p_\mu)v^*(p)  \  ,
\label{15}
\end{eqnarray}
where $\lambda^* $ is a constant, $m^*$ is the mass of the spin-1/2
state $|\Delta^*(p)\rangle $ and  $v^*(p )$
the relevant Dirac spinor satisfying $(\hat{p}-m^*)v^*(p)=0$.

{}From eq.(\ref{15}), it follows that any amplitude containing
the  $p^{\mu}$-factor, is ``contaminated'' by the transitions into
the spin-1/2 states.
These states lie higher than the $\Delta$-isobar and,
in principle, one can
treat them as a part of the continuum.
 Then, however, there will be  strong  reasons to expect that
the duality interval $S_0$ for the ``contaminated''
invariant amplitudes deviates from  that for the amplitudes
containing only the spin-3/2 contributions in the
$\eta_{\mu}$-channel.
Another possibility is to project out   the amplitudes
which  have  contributions due to the transitions into
 spin-$1/2$ isospin-$3/2$ states.
To achieve  this,
it is convenient to use  the basis
in which
 $\gamma_\mu$ is always placed at
the leftmost  position.
Then, according to  eqs.(\ref{5}-\ref{7}),
the  invariant amplitudes corresponding
to the structures  with  $q_\mu$ and $g_{\mu\nu}$
are free from  the contributions due to the
spin-1/2  isospin-3/2 states.
In this basis, keeping  only the
terms with $q_\mu$ and $g_{\mu\nu}$ in eq.(\ref{5}), we get
\begin{eqnarray}
T_{\mu\nu}^{\gamma^* p \to \Delta}(p,q) =
\frac{ l_N l_\Delta }{
(2\pi)^4
(p^2-M^2)((p-q)^2-m^2)} \nonumber
\\
\times\left(
g_{\mu\nu} [\hat{p},\hat{q}] \frac{3(M+m)}{8m}
(G_M^*(Q^2)+G_E^*(Q^2))\right.
 \nonumber \\
\left.
+\frac{q_\mu}{2} \left ( m [\gamma_\nu, \hat{p}]
+ M [\gamma_\nu, (\hat{p} - \hat{q})] \right ) G_1(Q^2)
\right.
 \nonumber \\
\left.
- p_\nu [\hat{p},\hat{q}] G_2(Q^2)
- q_\nu [\hat{p},\hat{q}] \left(G_3(Q^2) - \frac12 G_2(Q^2)
\right )+ ...\right)  \ .
\label{16}
\end{eqnarray}

Hence,   from the invariant amplitudes related to
the structures proportional to
$q_\mu$, we can get the
local duality estimates for the form factors
$G_1, G_2, G_3$. Similarly,
extracting the
structure $g_{\mu\nu} [\hat{p},\hat{q}]$, we  get
an expression for $(G_M^*+G_E^*)$.  Counting the powers of $q$,
we expect that results for $G_2$ are  less reliable than
those for $G_1$ and $G_M^*+G_E^*$, while results for $G_3$
are less reliable than those for $G_2$.

The number of independent amplitudes
can be diminished by
taking some  explicit projection of the
original amplitude   $T_{\mu\nu}(p,q)$.
 In particular, if one multiplies   $T_{\mu\nu}$ by $p_\nu$,
the invariant amplitude corresponding to the
structure $q_\mu [\hat{q},\hat{p}]$ is proportional to the
quadrupole form factor $G_C^*(Q^2)$:
\begin{eqnarray}
p_\nu T_{\mu\nu}^{\gamma^* p \to \Delta}(p,q)=
\frac{  l_N l_\Delta }
{(2\pi)^4 (p_1^2-m^2)(p_2^2 - M^2)}
\left  \{ \frac38\frac{M+m}{m} q_\mu [\hat{q},\hat{p}] G_C^*(Q^2)
+ \ldots \right \} \ .
\label{17}
\end{eqnarray}

Another possibility
is to take the trace of
$T_{\mu\nu}^{\gamma^* p \to \Delta}$. The result
 is proportional to
the magnetic form factor $G_M^*(Q^2)$:
\begin{eqnarray}
{\rm Tr}\, (T_{\mu\nu})=
\frac{  l_N l_\Delta }
{(2\pi)^4 (p_1^2-m^2)(p_2^2-M^2)}
\frac{M+m}{2m}
\left(
4 i\varepsilon^{\mu\nu\alpha\beta}q_\alpha p_\beta
\right) G_M^*(Q^2) \ .
\label{18}
\end{eqnarray}
However, one  should remember that the trace of $T_{\mu\nu}$
is not free from contributions due to  spin-1/2 isospin-3/2 states.

\section{Estimates for the $\gamma^* p \to \Delta$ form factors}

Though the invariant amplitude related to the trace of  $T_{\mu\nu}$
is contaminated by the transitions into spin-1/2 isospin-3/2 states,
it makes sense to consider  this amplitude
because it  has the simplest
perturbative spectral density:
\begin{eqnarray}
\frac1{\pi^2}\rho_M^{pert.}(s_1,s_2, Q^2)
=
\frac{Q^2}{8 \kappa^3}(\kappa-(s_1+s_2+Q^2))^2
(2\kappa+s_1+s_2+Q^2) \ ,
\label{19}
\end{eqnarray}
where
\begin{eqnarray}
\kappa=\sqrt{(s_1+s_2+Q^2)^2-4s_1s_2} \ .
\label{20}
\end{eqnarray}
Imposing  the   local duality prescription, we get
\begin{eqnarray}
G_M^*(Q^2)=\frac{2m}{ l_N l_\Delta (M+m)}
\int_0^{s_0}ds_1\int_0^{S_0} \frac{\rho_M^{pert.}(s_1,s_2,Q^2)}
{\pi^2} ds_2
 \nonumber
\\
=\frac{6m}{ (M+m)}F(s_0,S_0,Q^2) \  ,
\label{21}
\end{eqnarray}
where $F(s_0,S_0,Q^2)$ is a universal function
\begin{eqnarray}
F(s_0,S_0,Q^2)=  {{s_0^3 S_0^3 }\over
{9 l_N l_\Delta  (Q^2+s_0+S_0)^3}
\left ( 1-3 \sigma+(1-\sigma)\sqrt{1-4 \sigma} \right ) }
\label{22}
\end{eqnarray}
and  $\sigma = s_0S_0/(Q^2+s_0+S_0)^2$.
As we will see, the results for other invariant amplitudes
can be conveniently expressed through $F(s_0,S_0,Q^2)$.

The function $F(s_0,S_0,Q^2)$ depends on the duality
intervals $s_0$ and $S_0$.
We fix the nucleon duality interval $s_0$   at the
standard value  $s_0 = 2.3 \, GeV^2$ extracted from the analysis
of the two-point function and  used earlier in the
nucleon form factor calculations.
The results of the existing two-point function analysis
for the $\Delta$-isobar \cite{belioffe}
are compatible with  the $\Delta$ duality
interval $S_0$ in the range $3.2$ to $4.0 \, GeV^2$.
To fine-tune the $S_0$ value,
we consider two independent sum rules
for the $G_1$ form factor
\begin{eqnarray}
m G_1(Q^2)=
2 \left (3+ Q^2 \frac{d}{dQ^2} \right ) F(s_0,S_0,Q^2)
\nonumber
\\
-2 Q^2 \left( \frac{d}{dQ^2}\right)^2
\int_0^{S_0} F(s_0,s_2,Q^2) \, ds_2
\label{23}
\end{eqnarray}
and
\begin{eqnarray}
M G_1(Q^2) =
\frac{3}{2} Q^2\left(\frac{d}{dQ^2}\right)^2\int_0^{S_0}
F(s_0,s_2,Q^2) \, ds_2
\label{24}
\end{eqnarray}
extracted from the invariant amplitudes
corresponding to the structures
$q_\mu [\gamma_\nu ,\hat{p}]$ and
$q_\mu [\gamma_\nu ,(\hat{p} - \hat{q})]$,
respectively
(recall that $p-q$ is the proton's momentum
and $p$ is that of $\Delta$).
\begin{figure}[t]
  \epsfxsize=18cm
  \epsfysize=14cm

 \epsffile{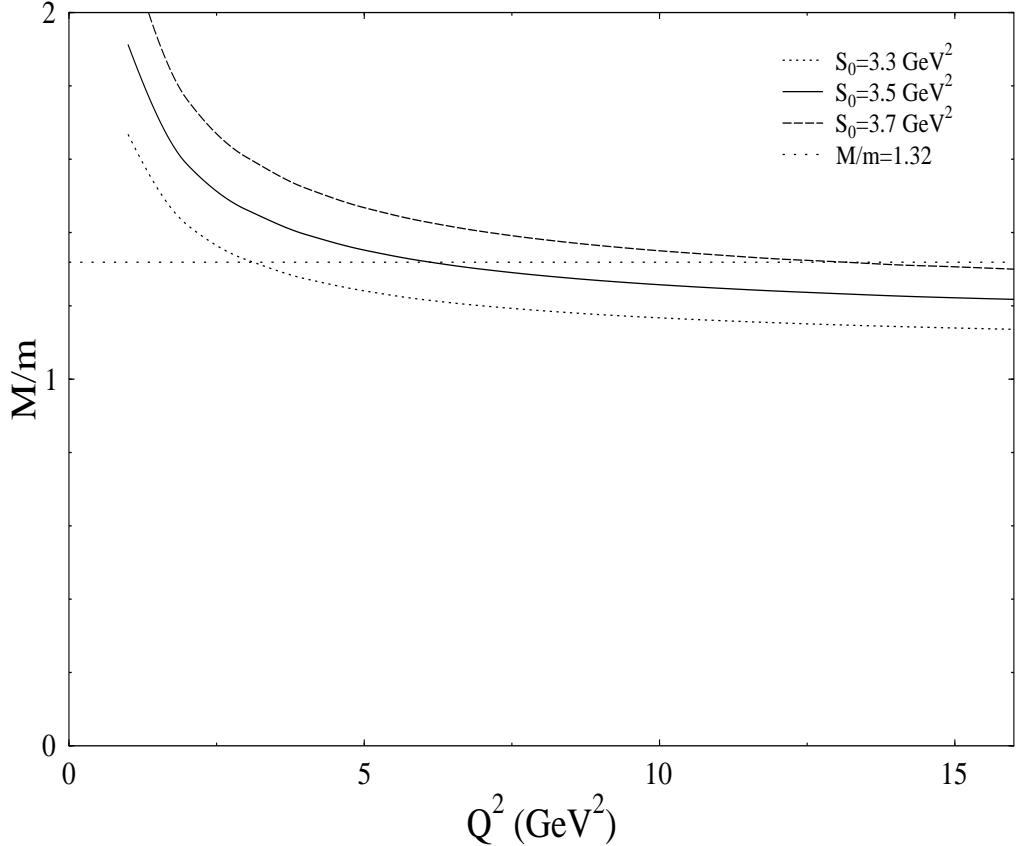}

 \vspace{-1cm}
 \centerline{\parbox{11cm}{\caption{\label{fig:1}
Isobar to  proton  mass ratio from the sum rules (24),(25).
   }}}
\end{figure}
Taking the ratio of these two  relations,
one can investigate  their mutual consistency and
 test the overall reliability of the
quark-hadron duality estimates.

Indeed, on the ``hadronic''  side, we have the ratio $M/m$
of the isobar and nucleon masses,
while on the ``quark'' side we have the ratio
of two explicit
and  non-trivially related functions.
The consistency requires, first, that the ratio
of these functions must be close to a constant and, second,
that this constant must be  close to the experimental value
for  the ratio of the isobar and nucleon masses:
$(M/m)^{exp} \approx 1.32$.
On Fig.1,  we plot the $Q^2$-dependence for the
ratio of the right hand sides
of eqs.(\ref{23}) and (\ref{24})
for the standard value $s_0 = 2.3 \, GeV^2$
of the nucleon duality interval
and three different values of $S_0$.
One can see that one should not rely on local duality estimates
below $Q^2 \sim 3 \, GeV^2$.
However, above  $Q^2 \sim 3 \, GeV^2$,
the ratio is pretty constant for all three values of $S_0$,
and rather close to 1.3.
The best  agreement is reached for $S_0=3.5 \, GeV^2$,
and we will use this value as the basic
isobar duality interval in further calculations.
In particular, the $l_{\Delta}$ parameter
will be fixed by
$l_{\Delta}^2 = \frac1{10}(3.5\, GeV^2)^3$
(cf. eq.(\ref{14})).

{}From
 eqs.(\ref{8}) and (\ref{9}), it follows that  $G_1$
 is proportional to the difference of
the magnetic $G_M^*$ and electric $G_E^*$ transition form factors:
\begin{equation}
G^{(-)}(Q^2) \equiv  G_M^*(Q^2) - G_E^*(Q^2) = \frac{2m}{3M(M+m)}
\left ( (M+m)^2 + Q^2 \right ) G_1(Q^2) \ .
\label{25}
\end{equation}

According to eq.(\ref{16}), the sum  $G^{(+)}(Q^2) \equiv G_M^*(Q^2) +
G_E^*(Q^2)$
of these form factors can be obtained
 from the  invariant
amplitude corresponding to
the  structure  $g_{\mu\nu}[\hat{p}, \hat{q}] $.
Applying the local duality prescription,
we obtain
\begin{eqnarray}
G^{(+)} (Q^2) = \frac{8m}{
M+m}\left[F(s_0,S_0,Q^2)
- \frac{Q^2}{12}\left(\frac{d}{dQ^2}\right)^2\int_0^{s_0}
F(s_1,S_0,Q^2) ds_1
\right ] \ .
\label{26}
\end{eqnarray}

Now, having expressions  both for $G^{(+)} (Q^2)$
 and $G^{(-)} (Q^2)$,
we can calculate $G_M^*(Q^2)$ and $G_E^*(Q^2)$.
The results for the combinations $Q^4 G_M^*(Q^2)$
and $ G_E^*(Q^2)/G_M^*(Q^2)$
are shown on Figs.2  and 3, respectively.
It should be  noted that, though  $F(s_0,S_0,Q^2)$ has the  $1/Q^6$
asymptotics for large $Q^2$ (see eq.(\ref{22}),
the local duality results are fairly
consistent with the $1/Q^4$-behaviour in the wide range
$5 \, GeV^2 \lapprox Q^2 \lapprox 20 \, GeV^2$.

\begin{figure}[t]

  \epsfxsize=18cm
  \epsfysize=14cm

 \epsffile{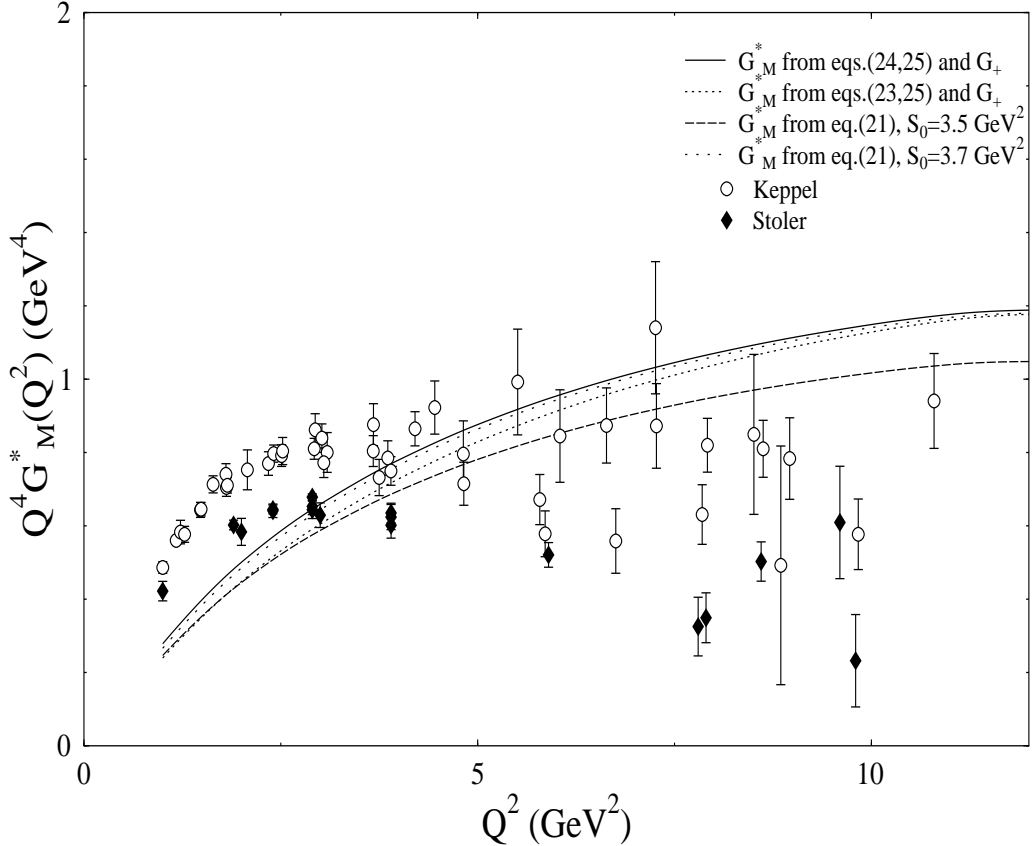}

 \vspace{-1cm}
 \centerline{\parbox{11cm}{\caption{\label{fig:2}
 Form factor $G_M^*(Q^2)$.
   }}}
\end{figure}

An important observation is that $G_E^*(Q^2)$ is
predicted to be much  smaller than
$G_M^*(Q^2)$ (see Fig.3).
It should be noted that if the $\gamma^* p \to \Delta$
transition form factors are calculated in a purely
pQCD approach (in which only the
$O((\alpha_s / \pi)^2 )$ double-gluon-exchange
diagrams are taken into account), the sum of  electric and
magnetic form factors
$G_M^*(Q^2)+G_E^*(Q^2)$ is
suppressed for asymptotically large $Q^2$
by a power of $1/Q^2$ \cite{carlson}.
This is because  the matrix element
\begin{eqnarray}
\langle 3/2|\Gamma|1,-1/2\rangle  \sim (G_M^*+G_E^*)
\label{27}
\end{eqnarray}
violates the helicity conservation requirement
for the hard subprocess amplitude.
In other words, the pQCD prediction is  that
 \mbox{$(G_M^*+G_E^*)$} should behave asymptotically like $1/Q^6$,
while each of $G_M^*$ and $G_E^*$ behaves like $1/Q^4$.
As a result, asymptotically $G_E^* \sim -G_M^*$.
However, we consider here only
the soft contribution  generated by the
Feynman mechanism for which
the helicity conservation arguments are not
applicable.
Thus,  for the soft term, there are no {\it a priori}
grounds to expect that $G_E^* \sim -G_M^*$.

\begin{figure}[t]

  \epsfxsize=18cm
  \epsfysize=14cm

 \epsffile{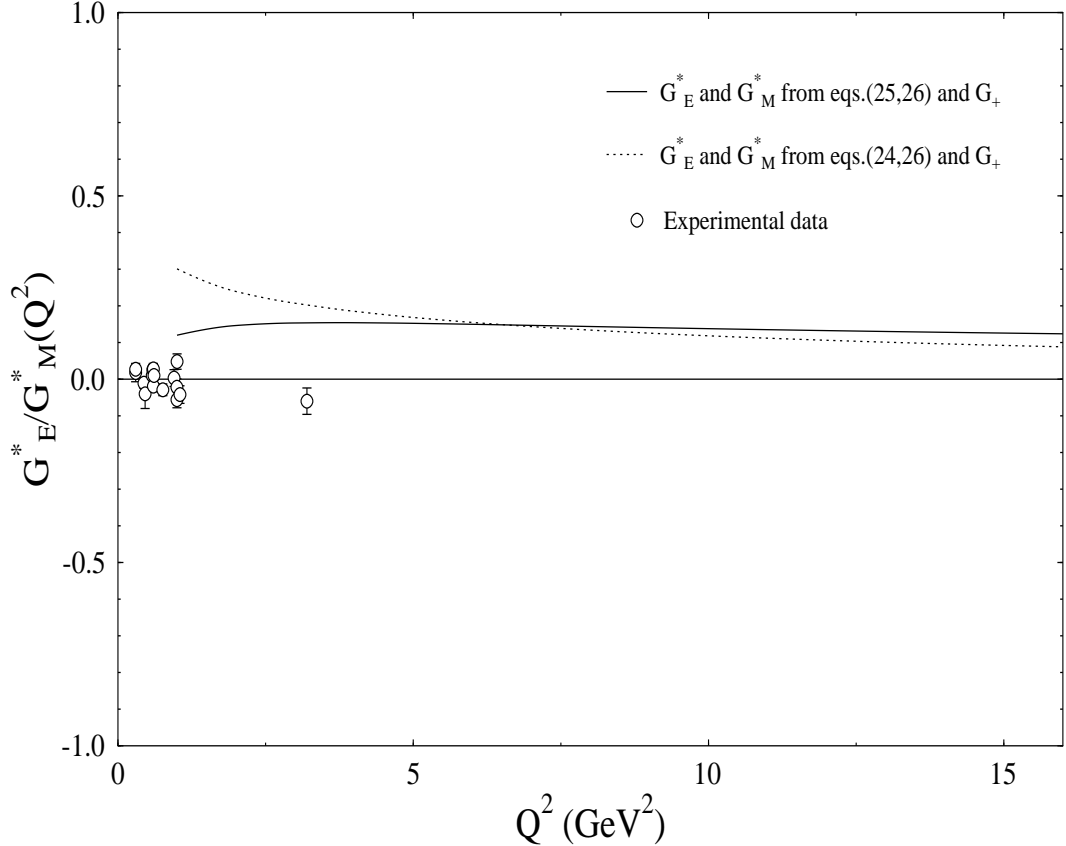}

 \vspace{-1cm}
 \centerline{\parbox{11cm}{\caption{\label{fig:3}
 Ratio of form factors $G_E^*(Q^2)$ and $G_M^*(Q^2)$ as
calculated from eqs. (24)-(27).  Experimental data
were taken from ref.[23] and the
point at $3.2 \, GeV^2$ from ref.[24].
   }}}
\end{figure}

The smallness of  $G_E^*(Q^2)/ G_M^*(Q^2)$
dictated by local duality, strongly contrasts  with the
pQCD-based expectation that $G_E^*(Q^2) \sim - G_M^*(Q^2)$,
and this allows for an experimental discrimination
between the two competing mechanisms.
One should  realize, however, that $G_E^*(Q^2)$
is  obtained in our calculation
as a small difference between
two large combinations $G^{(+)}(Q^2)$ and $G^{(-)}(Q^2)$,
both dominated by $G_M^*(Q^2)$.
Hence, even a relatively small uncertainty
in either of these combinations (which is always there,
since the local duality gives only
approximate estimates) can produce
a rather large relative uncertainty
in the values of $G_E^*$.
In this situation,  we  restrict
ourselves to a  conservative
statement  that the electric form factor $G_E^*(Q^2)$ is small
compared to $G_M^*(Q^2)$
in the whole experimentally accessible region
without insisting  on a specific
curve for $G_E^*(Q^2)$.

Experimental points for $G_M^*$ shown in Fig.2
were taken from  the results for the
$G_T(Q^2)$ form factor obtained from analysis
of inclusive  data \cite{stoler1},
\cite{keppel}.
Since our results give a very small value for
the ratio  $(G_E^*/G_M^*)^2$,
the $G_E^*$ term in eq.(\ref{GT}) can be neglected.
One can see that,
 in the  $Q^2 \gapprox 3 \, GeV^2$ region,
 the local duality predictions
$G_M^*(Q^2)$  are close to  the results
of the  recent  analysis \cite{keppel}.

The magnetic form factor $G_M^*(Q^2)$
can also be obtained from  eq.(\ref{21}).
If one takes  the basic  duality
interval $S_0=3.5 \, GeV^2$,
the resulting values of $G_M^*(Q^2)$ (Fig.2)
are  somewhat  smaller than
those obtained by combining the results for
$G^{(+)}(Q^2)$ and $G^{(-)}(Q^2)$.
As emphasized earlier,
the spin-1/2 states also contribute to the trace
of $T_{\mu\nu}$, and the duality interval in this
case can be different from the basic value.
In fact, taking $S_0 = 3.7 \, GeV^2$ in eq.(\ref{21}),
we get  a curve for $G_M^*(Q^2$ (Fig.2) essentially coinciding
with those  obtained from the sum of
 $G^{(+)}(Q^2)$ and $G^{(-)}(Q^2)$.

\begin{figure}[t]

  \epsfxsize=18cm
  \epsfysize=14cm

 \epsffile{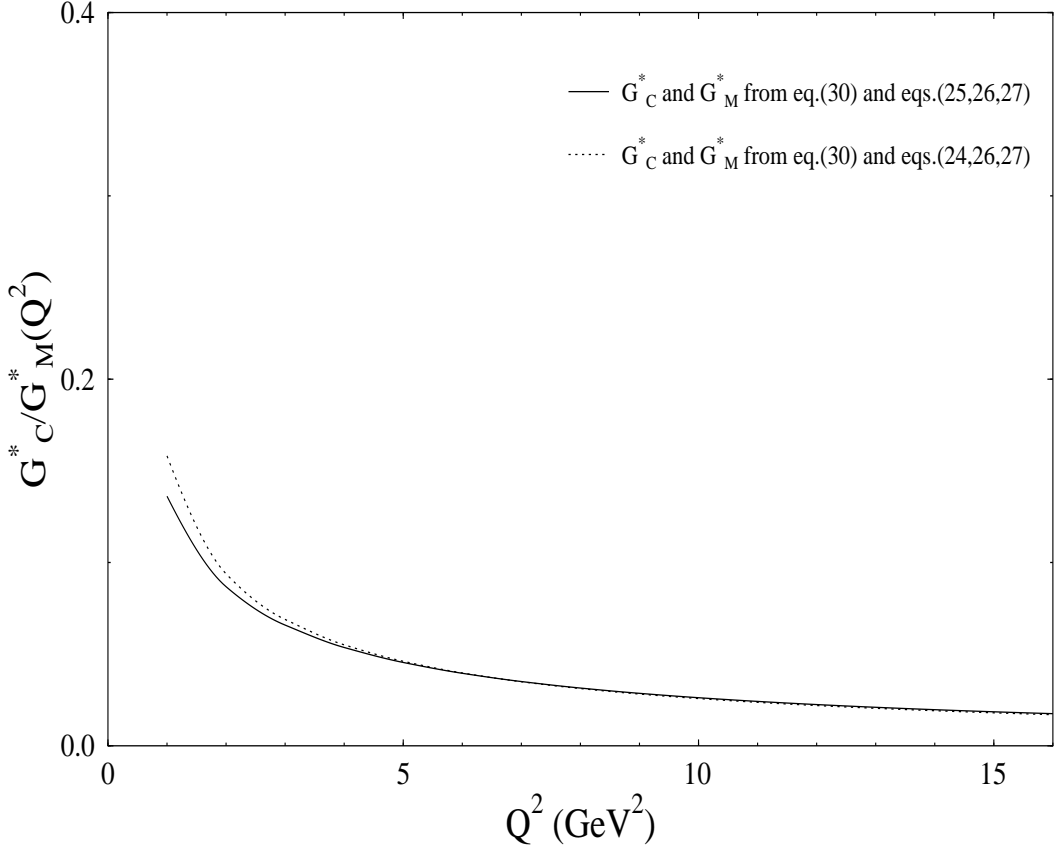}

 \vspace{-1cm}
 \centerline{\parbox{11cm}{\caption{\label{fig:4}
 Ratio  $G_C^*(Q^2)/G_M^*(Q^2)$ as calculated from eqs.
(24)-(27) and (30).
   }}}
\end{figure}

The  quadrupole
(Coulomb) form factor $G_C^*(Q^2)$
can  be calculated from the  expression  (\ref{17})
for the contracted amplitude $p_\nu T_{\mu\nu}$:
\begin{eqnarray}
G_C^*(Q^2)=\frac{8m}{3
(M+m)}\left[-\frac{d}{dQ^2}\int_0^{S_0} F(s_0,s_2,
Q^2) \, ds_2 \nonumber
\right.
\\
-\frac{Q^2}4\left(\frac{d}{dQ^2}\right)^2\int_0^{s_0}
F(s_1,S_0,Q^2) \,ds_1
\nonumber
\\
\left.+\frac12\left(\frac{d}{dQ^2}\right)^2
\left(1+\frac{d}{dQ^2}\right)\int_0^{s_0}ds_1\int_0^{S_0}
F(s_1,s_2,Q^2)\, ds_2 \right] \ .
\label{30}
\end{eqnarray}
Again, $G_C^*(Q^2)$  is essentially smaller than $G_M^*(Q^2)$
(see Fig.4).
Furthermore,  eq.(\ref{30}) predicts that, for large $Q^2$,
the quadrupole form factor $G_C^*(Q^2)$ has  an extra $1/Q^2$
suppression compared to $G_M^*(Q^2)$.
In fact, if the duality intervals
are  equal, $s_0 = S_0$, the suppression
  is even stronger, namely, by
two powers of $1/Q^2$.

\section{Conclusions}

We applied the local quark-hadron duality prescription
to  estimate  the soft contribution
to the $\gamma^* p \to \Delta $ transition form factors.
We observed a reasonable agreement  between the results
obtained from different invariant amplitudes.
We found that the transition is dominated by the magnetic form factor
$G_M^*(Q^2)$ while electric $G_E^*(Q^2)$
and quadrupole $G_C^*(Q^2)$ form factors are
small compared to $G_M^*(Q^2)$ for all experimentally accessible
momentum transfers. Numerically,
our estimates for $G_T(Q^2)$ are  close to
those obtained from a recent
analysis of inclusive  data \cite{keppel}.
Hence,  there is no need
for a sizable  hard-scattering contribution
to describe the data.
Furthermore, if future exclusive measurements  at CEBAF
would  show that
 the ratio $G_E^*(Q^2)/G_M^*(Q^2)$
is small above $Q^2 \sim 3 \, GeV^2$,
this would give an unambiguous experimental
 proof of the dominance
of the soft contribution.

\section{Acknowledgements}

We are very grateful to
P.Stoler, N.Isgur, V.Burkert and  C.E.Carlson  for
discussions which strongly motivated this investigation.
We thank C. Keppel
for providing us with the results of her analysis
and  F.Gross, R.Schiavilla,  M.J.Musolf,
W.J. van Orden, J.L. Goity
 and I.V.Musatov  for  interest in this work and critical comments.

This work was supported  by the US Department of Energy under
contract DE-AC05-84ER40150.

\end{document}